\documentclass[twoside,onecolumn,14pt]{article}
\usepackage{extsizes}
\usepackage[super,sort&compress,comma]{natbib} 
\usepackage[version=3]{mhchem}
\usepackage[left=1.5cm, right=1.5cm, top=1.785cm, bottom=2.0cm]{geometry}
\usepackage{balance}
\usepackage{times,mathptmx}
\usepackage{sectsty}
\usepackage{graphicx} 
\usepackage{lastpage}
\usepackage[format=plain,justification=justified,singlelinecheck=false,font=large,labelfont=bf,labelsep=space]{caption}
\usepackage{float}
\usepackage{fancyhdr}
\usepackage{fnpos}
\usepackage[english]{babel}
\addto{\captionsenglish}{%
  
}
\usepackage{array}
\usepackage{droidsans}
\usepackage{charter}
\usepackage[T1]{fontenc}
\usepackage[usenames,dvipsnames]{xcolor}
\usepackage{setspace}
\usepackage[compact]{titlesec}
\usepackage{hyperref}

\usepackage{geometry}                		
\usepackage{amsmath,amssymb,latexsym}
\usepackage{pifont}

\usepackage{xr}
\definecolor{cream}{RGB}{222,217,201}

\begin{document}

\makeFNbottom
\makeatletter
\renewcommand\LARGE{\@setfontsize\LARGE{15pt}{17}}
\renewcommand\Large{\@setfontsize\Large{12pt}{14}}
\renewcommand\large{\@setfontsize\large{10pt}{12}}
\renewcommand\footnotesize{\@setfontsize\footnotesize{7pt}{10}}
\makeatother

\renewcommand{\thefootnote}{\fnsymbol{footnote}}
\renewcommand\footnoterule{\vspace*{1pt}%
\color{cream}\hrule width 3.5in height 0.4pt \color{black}\vspace*{5pt}} 
\setcounter{secnumdepth}{5}

\makeatletter 
\renewcommand\@biblabel[1]{#1}            
\renewcommand\@makefntext[1]%
{\noindent\makebox[0pt][r]{\@thefnmark\,}#1}
\makeatother 
\renewcommand{\figurename}{\small{Figure}~}
\sectionfont{\sffamily\Large}
\subsectionfont{\sffamily\Large}
\subsubsectionfont{\bf}
\setstretch{1.125} 
\setlength{\skip\footins}{0.8cm}
\setlength{\footnotesep}{0.25cm}
\setlength{\jot}{10pt}
\titlespacing*{\section}{0pt}{4pt}{4pt}
\titlespacing*{\subsection}{0pt}{15pt}{1pt}

\newcommand{\beginsupplement}{%
        \setcounter{table}{0}
        \renewcommand{\thetable}{S\arabic{table}}%
        \setcounter{figure}{0}
        \renewcommand{\thefigure}{S\arabic{figure}}%
     }

\renewcommand*\rmdefault{bch}\normalfont\upshape
\rmfamily
\section*{}
\vspace{-1cm}

\sffamily

\vspace{0.3cm}
\noindent\textbf{Unravelling structural rearrangement of polymer colloidal crystals under dry sintering conditions} \\
\vspace{0.3cm}

\noindent\large{Alexey V. Zozulya,$^{\ast}$\textit{$^{a}$} Ivan A. Zaluzhnyy,\textit{$^{b, c}$} Nastasia Mukharamova,\textit{$^{b}$} Sergey Lazarev,\textit{$^{b,d}$}
 Janne-Mieke Meijer,\textit{$^{e\ddag}$} Ruslan P. Kurta,\textit{$^{a}$} Anatoly Shabalin,\textit{$^{b\S}$} Michael Sprung,\textit{$^{b}$} 
 Andrei V. Petukhov,\textit{$^{e, f}$} and Ivan A. Vartanyants\textit{$^{b, c}$}} 


\begin{abstract}
\noindent\large{
Structural rearrangement of polystyrene colloidal crystals under dry sintering conditions has been revealed by \textit{in situ} grazing incidence X-ray scattering. Measured diffraction patterns were analysed using distorted wave Born approximation (DWBA) theory and the structural parameters of as-grown colloidal crystals of three different particle sizes were determined for in-plane and out-of-plane directions in a film. By analysing the temperature evolution of diffraction peak positions, integrated intensities, and widths the detailed scenario of structural rearrangement of crystalline domains at a nanoscale has been revealed, including thermal expansion, particle shape transformation and crystal amorphisation. Based on DWBA analysis we demonstrate that in the process of dry sintering the shape of colloidal particles in a crystal transforms from a sphere to a polyhedron. Our results deepen the understanding of thermal annealing of polymer colloidal crystals as an efficient route to the design of new nano-materials.}\\
\end{abstract}

\footnotetext{\textit{$^{a}$~European XFEL GmbH, Holzkoppel 4, D-22869 Schenefeld, Germany; E-mail: alexey.zozulya@xfel.eu}}
\footnotetext{\textit{$^{b}$~Deutsches Elektronen-Synchrotron DESY, Notkestra\ss e 85, D-22607 Hamburg, Germany }}
\footnotetext{\textit{$^{c}$~National Research Nuclear University MEPhI (Moscow Engineering Physics Institute), Kashirskoye ch. 31, 115409 Moscow, Russia}}
\footnotetext{\textit{$^{d}$~National Research Tomsk Polytechnic University (TPU), Lenin Avenue 30, 634050 Tomsk, Russia}}
\footnotetext{\textit{$^{e}$~Van 't Hoff Laboratory for Physical and Colloid Chemistry, Department of Chemistry and Debye Institute for Nanomaterials Science, Utrecht University, Padualaan 8, 3584 CH, The Netherlands}}
\footnotetext{\textit{$^{f}$~Laboratory of Physical Chemistry, Department of Chemical Engineering and Chemistry and Institute for Complex Molecular Systems, Eindhoven University of Technology, P.O. Box 513, 5600 MB, Eindhoven, The Netherlands}}
\vspace{1.0cm}

\section{Introduction}
Photonic crystals are optical structures with periodic modulations of medium dielectric susceptibility leading to the appearance of photonic band gaps \cite{PhotonicCrystals}. Due to their optical properties of confining and controlling electromagnetic waves in three dimensions, photonic crystals have found various applications in optoelectronics \cite{Chow}, acousto-optics \cite{Gorishnyy}, photovoltaics \cite{Wehrspohn} and thermoelectrics \cite{Nutz2015}. Photonic crystals can be fabricated in a highly controllable fashion by nanolithography methods \cite{Campbell,Subramania}, which involve many-step fabrication procedure and are quite expensive. The self-assembly of colloidal particles into 3D crystals provides an alternative fabrication technique to nanolithography, with extended possibilities of tailoring optical properties of a photonic crystal by particle size \cite{Stimulak}. In this way, colloidal crystals can be synthesized from a colloidal suspension using convective assembly methods under ambient conditions \cite{Hilhorst, Meijer}, representing a cost-efficient and tunable fabrication technique. 

The shape of colloidal particles is responsible for the self-assembly of colloidal crystals \cite{Damasceno_ACSNano,Damasceno,Petukhov2015,Petukhov2017}, and directly affects their dimensionality \cite{OBrien}, phase behaviour \cite{Agarwal,Meijer2017} and optical properties \cite{Gong}. Thermal or chemical annealing can be used to modify the shape of self-assembled colloidal particles and, hence, tailor the photonic band gap properties of a colloidal crystal \cite{Miguez,Gates,Kuai,Liu2014,Nutz2017}. As a result of annealing treatment, the spherical shape of a colloidal particle changes to a faceted shape, which provides a new route for the fabrication of polyhedral particles \cite{Vutukuri}. Therefore, the pathway of particle shape transformation under thermal or chemical annealing conditions is an important aspect in structural studies of colloids, including potential applications of polyhedral colloidal particles for the synthesis of reconfigurable colloidal materials \cite{Du} and in the catalysis research \cite{Tian,Becknell}. 

Different methods can be employed to characterise colloidal systems depending on the size of particles. Optical and confocal laser scanning microscopies can be efficiently applied to visualize 3D colloidal systems \cite{Schall}, albeit the resolution of optical methods is limited to about 500 nm. Smaller than 500 nm particles can be viewed by scanning electron microscopy (SEM) or transmission electron microscopy (TEM) at sub-nanometer resolution \cite{Hatton}. However, due to low penetration depth of electrons, SEM and TEM methods can probe mostly the surface morphology of a specimen at a limited field of view. X-ray scattering methods, being free from these limitations due to high penetration depth, enable a non-destructive access to the interior of colloidal crystals at nanometer resolution and below. Furthermore, X-ray methods can probe sample areas ranging from hundreds of microns down to tens of nanometers, thus providing both mesoscopic and nanoscale structural information. Finally, X-ray techniques are well suited for \textit{in situ} measurements involving complex sample environments (e.g. temperature \cite{Zozulya}, pressure \cite{MSchroer}, chemical conditions \cite{Murphy}). 

Synchrotron X-ray scattering has been extensively applied for structural studies of colloidal crystals at ambient conditions \cite{PetukhovPRL, PetukhovJAC, Thijssen, Roth, Huber, Gulden, Shabalin, Lazarev}. Rearrangement of polymer colloidal assemblies under thermal annealing conditions has been studied by \textit{in situ} small angle X-ray scattering (SAXS) \cite{Hu2009, Hu2010, Chen, Herzog}. Recently, the details of structural transitions in polystyrene (PS) colloidal crystals upon heating treatment, including particle faceting and crystal melting, have been revealed \cite{Zozulya, Sulyanova}. In the present work, we employ \textit{in situ} grazing incidence X-ray scattering technique to unravel the details of structural evolution of PS colloidal crystal domains under dry sintering conditions. As compared to standard transmission SAXS geometry, grazing incidence X-ray scattering is sensitive to both in-plane and out-of-plane structural order, and furthermore provides better control during thermal annealing.

\section{Methods}
\subsection{Experiment}
\begin{figure}
	\includegraphics[bb=0 0 1200 600, width=\linewidth]{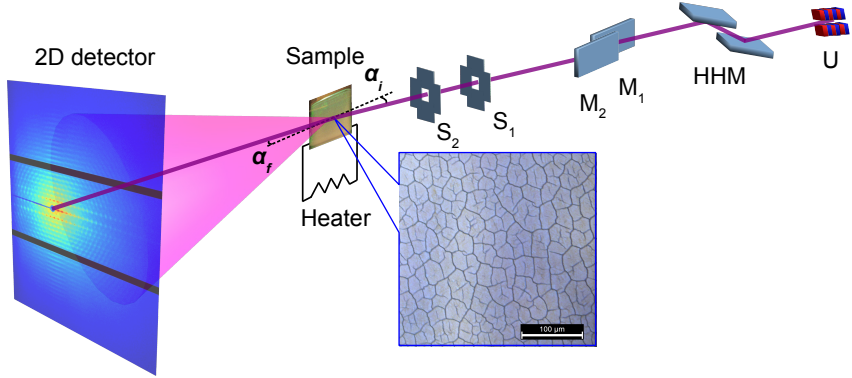}
	\centering
	\caption{Schematics of the experimental setup. X-ray beam produced by an undulator (U) is guided through a high-heatload monochromator (HHM) and two X-ray mirrors ($M_{1}$, $M_{2}$). The size of the beam impinging on a sample at grazing angle $\alpha_{i}$ was adjusted by the beam-defining slits $S_{1}$ and $S_{2}$. The sample was mounted on a heater stage inside the vacuum chamber and the scattering signal was acquired using the 2D detector placed downstream from the sample. Inset: optical micrograph of a PS colloidal crystal sample; scale bar 100 $\mu$m.}
	\label{img:Figure1_setup}
\end{figure}

Schematics and details of the experimental setup are presented in Fig. \ref{img:Figure1_setup}. X-ray scattering experiments were performed at the P10 coherence beamline of PETRA III synchrotron source at DESY, Hamburg. Incident X-ray beam with a photon energy of 7.74 keV was adjusted to 50$\times$50 $\mu$m$^2$ size using beam-defining slits. The sample was aligned vertically on a 5-axis positioning stage such that the scattering plane was horizontal. Diffraction patterns were acquired using 2D detector Pilatus 300K with an area of 487$\times$619 pixels and a pixel size of 172 $\mu$m$^2$. The detector was positioned at 5.1 m distance downstream the sample. The incident angle $\alpha_{i}$ for the studied samples was adjusted to 0.5$^{\circ}$. The scattering patterns were acquired by collecting 100 detector frames with an exposure time of 1 s each. Our measurements benefited from high incident flux of synchrotron X-ray beam of 1.1$\times$10$^{11}$ photons/s/50$\times$50 $\mu$m$^2$ and the fact that colloidal crystals were grown on thin glass coverslips of 170 $\mu$m thickness, providing relatively high transmission of 22\% at the used photon energy. The air scattering background was eliminated by positioning the sample inside of a vacuum chamber and using an evacuated flight tube between sample and detector. To protect the detector from a transmitted beam and optimize the scattering signal acquisition at high q-values the beamstop (tungsten cylinder of 3 mm diameter) was positioned inside of the flight tube in front of the detector.

\subsection{Sample Preparation}
PS colloidal crystal films were fabricated by vertical deposition method \cite{Meijer}. Spherical colloidal particles were synthesised by polymerization of an aqueous solution of styrene using potassium persulphate as initiator. Initial colloidal suspensions for the studied samples with particle diameter of 420, 272 and 194 nm (referred further as samples A, B and C, respectively) contained 0.8, 0.25 and 0.5\% volume fraction of colloidal particles in water. Particle polydispersity values measured by dynamic light scattering (DLS) were 2.1, 4 and 3.5\% (see Table \ref{Table1}). Thin glass substrates were immersed into colloidal suspensions and subsequently dried at 50$^\circ$C for several days at ambient conditions. The fabricated colloidal crystal films consisted of 20-80 monolayers of PS spherical particles and typically exhibited a cracked texture composed of domains with an average size of 10 - 50 $\mu$m (typical optical microscopy image is shown as insert in Fig. \ref{img:Figure1_setup}). The colloidal crystal samples were mounted to the sample holder using silver paste to assure a good thermal contact. The copper block of the sample holder was integrated into the vacuum chamber. Heating was supplied by two parallel connected heating elements integrated into the copper block, and the temperature was measured using two PT100 sensors embedded into the sample holder. Temperature and heating power were controlled using LakeShore 340 temperature controller. During the measurements the temperature of a sample was raised incrementally starting from room temperature (RT). After each temperature increment, a waiting time of 5 min. was applied before collecting the data to enable any kinetic processes within the sample to be completed.

\section{Results and discussion}
\subsection{GTSAXS simulations}
\begin{figure}
	\centering
	\includegraphics[bb=0 0 1200 600, width=\linewidth]{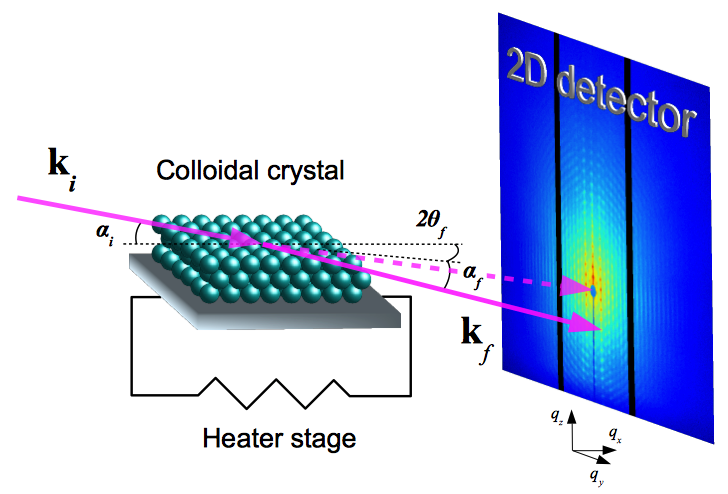}
	\caption{Scattering geometry of GTSAXS experiment. The incident beam with a wave vector \textbf{k$_i$} enters the sample surface at a grazing angle $\alpha_{i}$. The scattered beam with a wave vector \textbf{k$_f$} is defined by the in-plane and out-of-plane scattering angles $2\theta_{f}$ and $\alpha_{f}$, respectively. }
	\label{img:Figure1}
\end{figure}

To elucidate the discussion of the results we first consider the geometry of X-ray scattering experiment under grazing incidence conditions. In a conventional grazing incidence small angle X-ray scattering (GISAXS) geometry \cite{Renaud, Hexemer} the incident beam with a wave vector \textbf{k$_{i}$} illuminates the sample surface at a grazing angle $\alpha_{i}$, typically exceeding the critical angle of total external reflection $\alpha_c$ of the substrate material, and generates the scattered beam with a wave vector \textbf{k$_{f}$} exiting the surface at an angle $\alpha_{f}$. The direction of the scattered wave vector is defined by two angles, $2\theta_{f}$ and $\alpha_{f}$, for the in-plane and out-of-plane directions (Figure \ref{img:Figure1}). The components of the wave vector transfer \textbf{q}=\textbf{k$_f$}-\textbf{k$_i$} are related to the angular coordinates:
\begin{equation}
q_x=k_{0}[\cos(2\theta_{f})\cos(\alpha_{f})-\cos(\alpha_{i})],
\end{equation}
\begin{equation}
q_y=k_{0}\sin(2\theta_{f})\cos(\alpha_{f}),
\end{equation}
\begin{equation}
q_z=k_{0}[\sin(\alpha_{f})+\sin(\alpha_{i})].
\end{equation}

For the colloidal crystals studied in this work, the specularly reflected signal was hampered by surface roughness, that prevented us from using GISAXS geometry. Instead, we have employed grazing incidence transmission small angle X-ray scattering (GTSAXS) geometry \cite{Lu,Mahadevapuram}, where only the forward transmitted part of a scattering pattern below the sample horizon ($\alpha_{f}$<0) is analysed, as illustrated in Figure \ref{img:Figure1}.

In order to determine the structural parameters of colloidal samples the GTSAXS data were analysed by X-ray scattering simulations in the frame of DWBA theory \cite{Sinha, Rauscher, Lee, Tate, Zozulya2007}. In the case of colloidal particles  supported on a substrate, DWBA implies to consider the scattering amplitudes in the particle and at the interface between vacuum and substrate, as described by Fresnel reflection and transmission coefficients. According to DWBA, in total four scattering amplitudes contribute to the resulting scattering signal from a single particle: (1) the amplitude of kinematical scattering from a particle described by Born approximation, (2) the reflection at an interface followed by the scattering event in a particle, (3) the scattering in a particle followed by the reflection at an interface and (4) the reflection at an interface followed by the scattering event and subsequent reflection at an interface. In GTSAXS geometry only the first two scattering channels contribute to the resulting scattering amplitude of a single particle $A_{p}(\textbf{q})$=$A_{1}(\textbf{q})$+$A_{2}(\textbf{q})$:

\begin{equation}
A_1(\textbf{q}) = \chi_{0}(k^2/4\pi)\int S(\textbf{r}) e^{-i(\textbf{q}_{||}+\textbf{q}_{z,1})\textbf{r}}d\textbf{r},
\end{equation}
\begin{equation}
A_2(\textbf{q}) = \chi_{0}(k^2/4\pi)R(\alpha_{i})\int S(\textbf{r}) e^{-i(\textbf{q}_{||}+\textbf{q}_{z,2})\textbf{r}}d\textbf{r},
\end{equation}
where $\textbf{q}_{||}=(q_{x},q_{y})$ represent in-plane components and $q_{z,1}=-k_{z}^{f}-k_{z}^{i}$, $q_{z,2}=-k_{z}^{f}+k_{z}^{i}$ are out-of-plane components of the wave vector transfer. $k=2\pi/\lambda$ is the wave vector in vacuum and $\chi_{0}$ is the 0-th order Fourier component of susceptibility of the particle material at a given X-ray wavelength $\lambda$. The shape function $S(\textbf{r})$ is equal to unity inside of a particle and zero outside. $R(\alpha_{i})=(k_{z}^{i}-\widetilde{k_{z}}^{i})/(k_{z}^{i}+\widetilde{k_{z}}^{i})$ is Fresnel reflection coefficient for the incident wave. $k_{z}$ and $\widetilde{k_{z}}$ are z-components of the wave vector in vacuum and substrate media, respectively. The total scattering amplitude $A_{cr}(\textbf{q})$ from a colloidal crystal lattice consisting of $N$ monodisperse particles is the sum of scattering amplitudes from each particle with a radius-vector $\textbf{r}_{i}$:

\begin{equation}
A_{cr}(\textbf{q}) = A_{p}(\textbf{q}) \sum_{i=1}^{N}e^{-i\textbf{q}\textbf{r}_{i}}.
\end{equation}
The GTSAXS intensity distribution has been calculated as $I_{GTSAXS}(\textbf{q})$=$|A_{cr}(\textbf{q})|^{2}$.

\subsection{GTSAXS analysis}
\begin{figure}
	\centering
	\includegraphics[bb=0 0 1200 800, width=\textwidth]{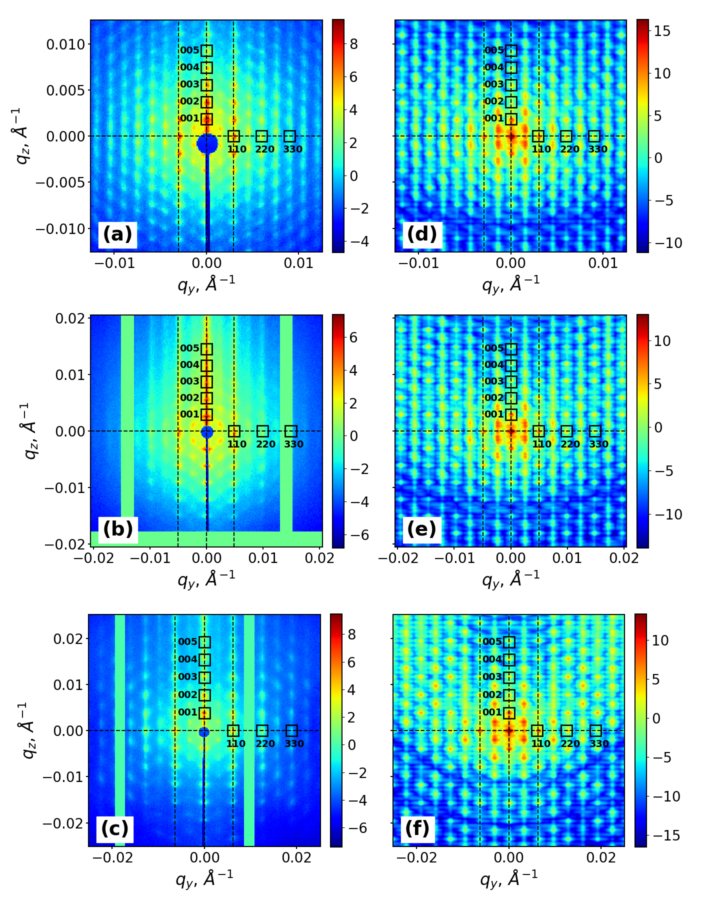}
	\caption{(a--c) Experimental and (d--f) simulated GTSAXS patterns for the three PS colloidal crystals (a,d) A, (b,e) B and (c,f) C at RT. Indexed sets of in-plane and out-of-plane reflections are marked by rectangles. Dashed lines indicate the directions of $q_{y}$- and $q_{z}$-profiles used for the analysis. A circular area around the direct beam was covered by the beamstop, and the vertical stripes of missing data are due to the gaps between detector tiles; horizontal stripe in  (b) was not captured by the detector. Intensities are represented in logarithmic scale.} 
	\label{img:Figure2}
\end{figure}

GTSAXS patterns measured at RT for the three colloidal crystals A, B and C, consisting of PS spherical particles of 420, 272 and 194 nm in diameter, are shown in Figure \ref{img:Figure2}a--c. Scattering patterns are displayed for the reciprocal space areas covering the same number of diffraction peaks, which corresponds to q-ranges 0.025$\times$0.025, 0.041$\times$0.041 and 0.050$\times$0.050 $\text{\AA}$$^{-1}$ for the samples A, B and C, respectively. The achieved resolution in reciprocal space of 1.32$\times$10$^{-4}$ $\text{\AA}$$^{-1}$ was defined by the pixel size of the 2D detector and the sample-to-detector distance at 7.74 keV X-ray energy (see Methods section for details). Indexed sets of in-plane reflections of (110)-type and out-of-plane reflections of (001)-type employed for the analysis are depicted by rectangles. Dashed lines denote the $q_{y}$- and $q_{z}$-directions used to extract intensity profiles for fitting analysis. In the case of GTSAXS geometry the scattering pattern is dominated by the kinematical scattering contribution and less influenced by re-scattering channels inherent to conventional GISAXS measurement. In particular, the characteristic GISAXS features such as specular and Yoneda peaks are not present in GTSAXS data. For all measured samples the central and side peaks along the $q_{z}$-direction reveal strong truncation rods which originate from the substrate scattering under grazing incidence angles. Due to variations of particle polydispersity and defect concentration we observed different number of diffraction orders for the studied colloidal crystals. For the sample A  the diffraction peaks up to 13-th order were observed indicating a high structural quality. Meanwhile, for the sample B only 5 diffraction orders have been observed pointing to rather low quality of the sample, which can be related to high polydispersity of 4\% (see Table \ref{Table1}). The sample C has revealed 7 diffraction orders, indicating an intermediate degree of long-range ordering. 

To obtain quantitative sample characteristics, GTSAXS simulations using Eqs. (1-6) were implemented as in-house developed Python scripts. Simulated patterns were calculated on 1600$\times$1600 px$^{2}$ grid for the q-ranges corresponding to the \mbox{GTSAXS} data from three PS colloidal crystals (Figure \ref{img:Figure2}d--f). To simulate the total scattering amplitude from a colloidal crystal we first calculated the scattering amplitude from a single particle according to DWBA, which was then summed over the particles constituting a single layer and, lastly, the summation over a stack of particle layers was performed. The stacking of $N_{l}$ particle layers in random hexagonal close-packed ($rhcp$) arrangement \cite{Meijer_JAC} was implemented by introducing $n_{hcp}$ hexagonal close-packed $hcp$ layers (layer sequence ABAB...) and $(N_{l}-n_{hcp})$ layers of face-centered cubic ($fcc$) structure (layer sequence ABCABC...). Peak broadening caused by lattice spacing variation was taken into account by a convolution of the simulated data with a Voigt function, which has also included the instrumental broadening due to the incident beam divergence of 4$\times$28 $\mu$rad$^2$ (vertical$\times$ horizontal) \cite{Zozulya}. 

\begin{table}[h]
\small
  \caption{\ Structural parameters of PS colloidal crystal samples at RT}
  \label{Table1}
  \begin{tabular*}{0.48\textwidth}{@{\extracolsep{\fill}}llccc}
    \hline
    Sample   &A                 &B                &C \\
    \hline
    $D_{DLS}$, $nm$                                    & 420          & 272          & 194  \\           
    $D_{GTSAXS}$, $nm$                            & 419          & 257          & 197  \\
    Polydispersity, \%                                    & 2.1             & 4             & 3.5    \\
    $c/a$ ratio                                               & 0.99         & 1.00         & 0.98 \\
    Packing                                                   & $rhcp$     & $rhcp$    & $rhcp$ \\
    $hcp$ fraction                                         & 0.6            & 0.2            & 0.8 \\
    $fcc$ fraction                                           & 0.4            & 0.8            & 0.2 \\
    Number of layers                                     & 30             & 70             & 60 \\
    Thickness, $\mu$$m$                             & 10.3          & 14.7          & 9.7 \\
    Lateral domain size, $\mu$$m$              & 4.2            & 2.3           & 3.7  \\
    In-plane $\Delta$$d/d$, \%                      & 4.3            & 5.4           &  4.8 \\
    Out-of-plane $\Delta$$d/d$, \%               & 5.1          & 6.8         & 3.5 \\
    In-plane mosaicity, $mrad$                      & 51           & 68           &  35 \\
    Out-of-plane mosaicity, $mrad$               & 43           & 54          &  48 \\
    \hline
  \end{tabular*}
\end{table}

The experimental $q_{y}$- and $q_{z}$-profiles extracted from GTSAXS patterns along the directions depicted by dashed lines in Figure \ref{img:Figure2} were fitted with calculated curves by adjusting the structural parameters such as particle diameter, lateral domain size, number of layers, lattice deformation and mosaicity values. Thus obtained best-fit curves (Supplementary Figure \ref{img:S1}) have yielded the in-plane and out-of-plane structural parameters of studied colloidal crystals at RT conditions, which are summarised in Table \ref{Table1}. Particle diameter values in the colloidal crystals determined by GTSAXS slightly deviate (within several per cents) from the results of DLS measurements on colloidal suspensions. Furthermore, for the samples A and C we found that the in-plane lattice spacing \textit{a} is by 1-2\% larger than the out-of-plane spacing \textit{c}. The observed in-plane lattice dilataion might be caused by $rhcp$ stacking nature of particle layers as well as surface tension affecting the formation of a colloidal crystal during solvent evaporation. By varying the $hcp$ and $fcc$ fractions in $rhcp$ stacking the corresponding values were evaluated within 10\% accuracy. Using the determined values of particle diameter and number of layers the total thickness of a colloidal crystal film can be estimated, which was found to increase from 9.7 $\mu$m for the sample C up to 14.6 $\mu$m for the sample B, with an intermediate value of 10.3 $\mu$m for the sample A. The determined values of lateral domain size, lattice deformation and mosaicity indicated a highest degree of structural quality of the sample A, a lowest quality of the sample B and an intermediate degree for the sample C.

\subsection{Temperature evolution of GTSAXS peak parameters}
\begin{figure}
	\centering
	\includegraphics[ bb=0 0 1250 750, width=\textwidth ]{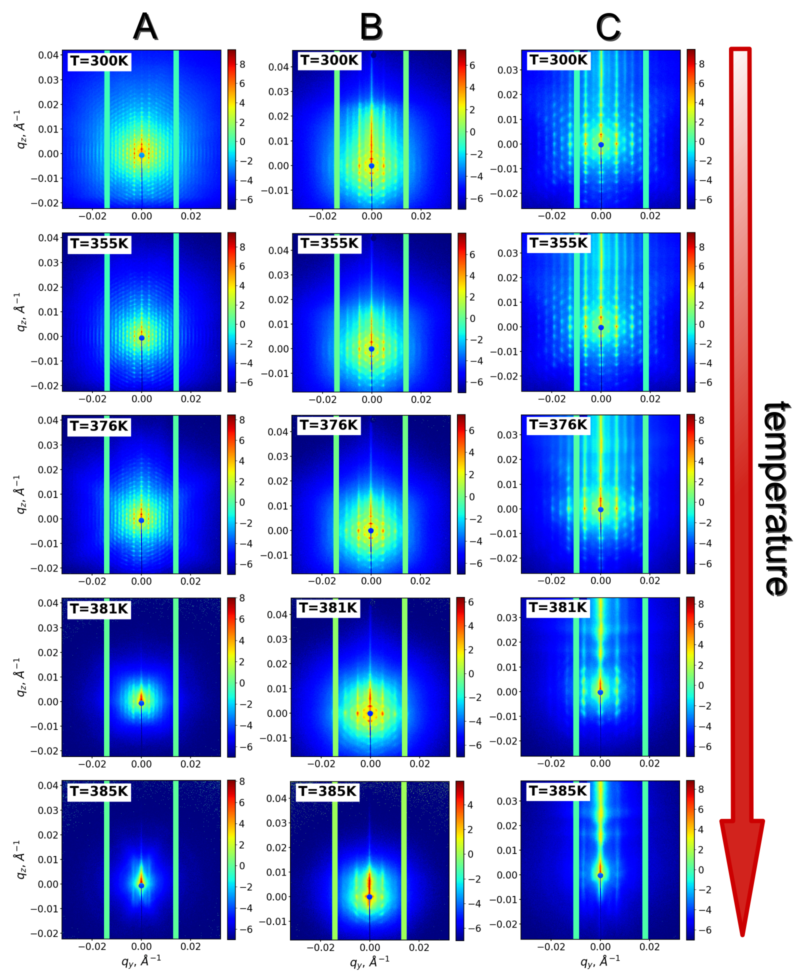}
	\caption{\textit{In situ} GTSAXS patterns for the PS colloidal crystal samples A (left column), B (middle column) and C (right column) measured during incremental heating in the range from RT to $T$ = 385 K. Intensities are represented in logarithmic scale.}
	\label{img:Figure3}
\end{figure}

Once the structural parameters of the colloidal crystals were quantified at RT conditions, we analysed \textit{in situ} GTSAXS data measured upon incremental heating of the samples in a wide temperature range from RT to 385 K. Figure \ref{img:Figure3} shows GTSAXS patterns for the samples A, B and C at selected temperatures: 300, 355, 376, 381 and 385 K. The patterns are displayed for the q-range of 0.0643$\times$0.0643 $\text{\AA}$$^{-1}$ representing almost full range of the 2D detector (square area of 487$\times$487 px$^2$ of Pilatus 300K detector, see Methods section). By visual inspection of diffraction peaks and diffuse scattering distribution one can readily identify the intermediate states of a colloidal crystal in the process of dry sintering. In the temperature range from RT to annealing temperature $T_{a}$ = 355 K the peak shapes and positions exhibit no significant changes, while the isotropic diffuse scattering becomes suppressed. With a further increase of temperature up to $T$ = 376 K the high-order diffraction peaks start to vanish, indicating a gradual loss of long-range ordering in a colloidal crystal. The fading of long-range order is accompanied by a transition from isotropic to anisotropic diffuse scattering distribution in the form of two inclined flares (see Figure \ref{img:Figure3} at $T$ = 376 K). The anisotropy of diffuse scattering indicates the onset of particle plastic deformation and shape transformation from a spherical to a faceted shape. Above $T$ = 381 K the abrupt decay of lower diffraction orders signifies the total loss of ordering and the formation of an amorphous polymer film.

\begin{figure}
	\centering
	\includegraphics[bb=0 0 3500 3500, width=\linewidth ]{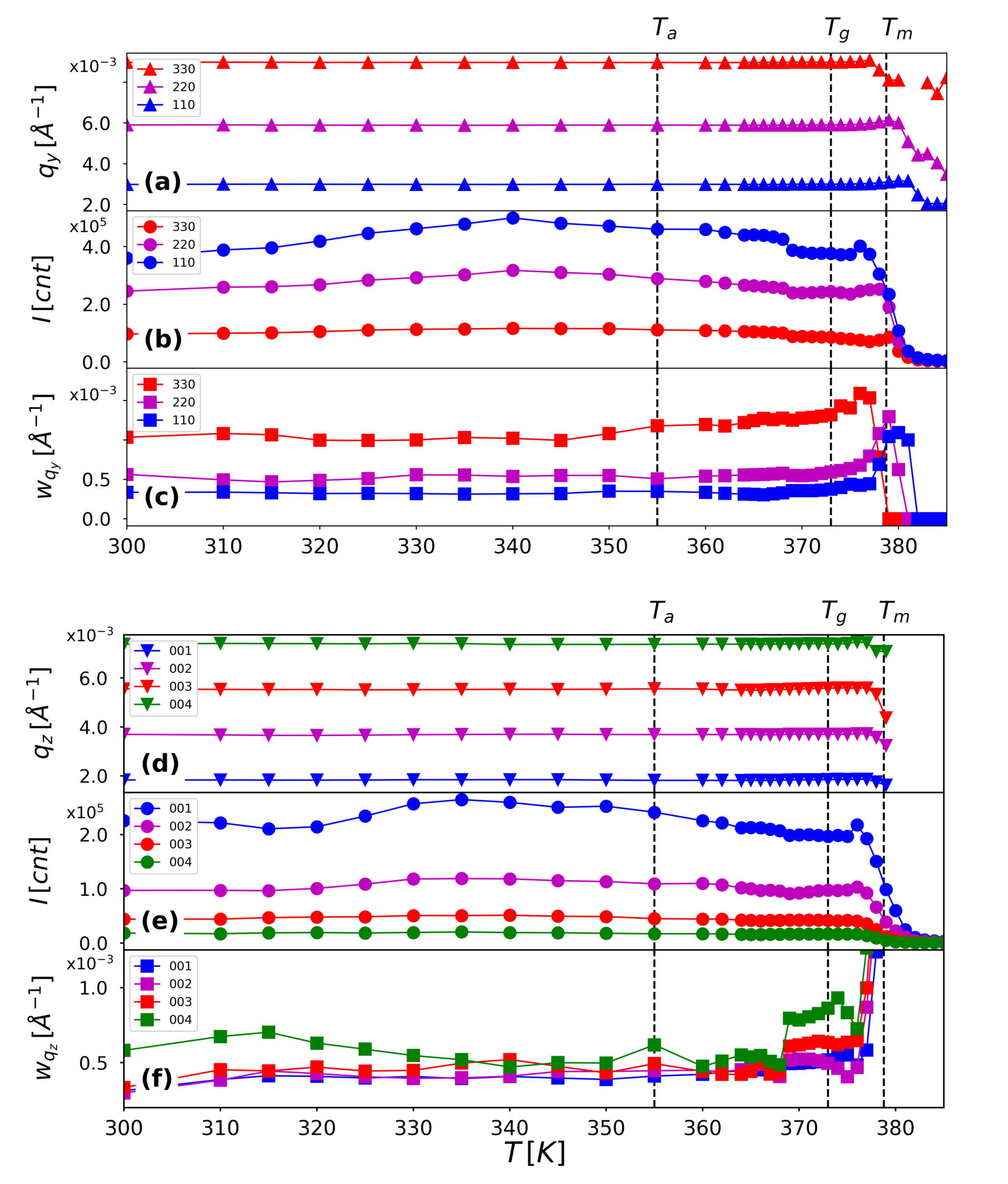}
	\caption{Temperature dependences of GTSAXS peak positions, integrated peak intensities and widths of the PS colloidal crystal sample A for (a--c) in-plane and (d--f) out-of-plane directions.}
	\label{img:Figure4}
\end{figure}

\begin{figure}
	\centering
	\includegraphics[bb=0 0 1200 600, width=\columnwidth]{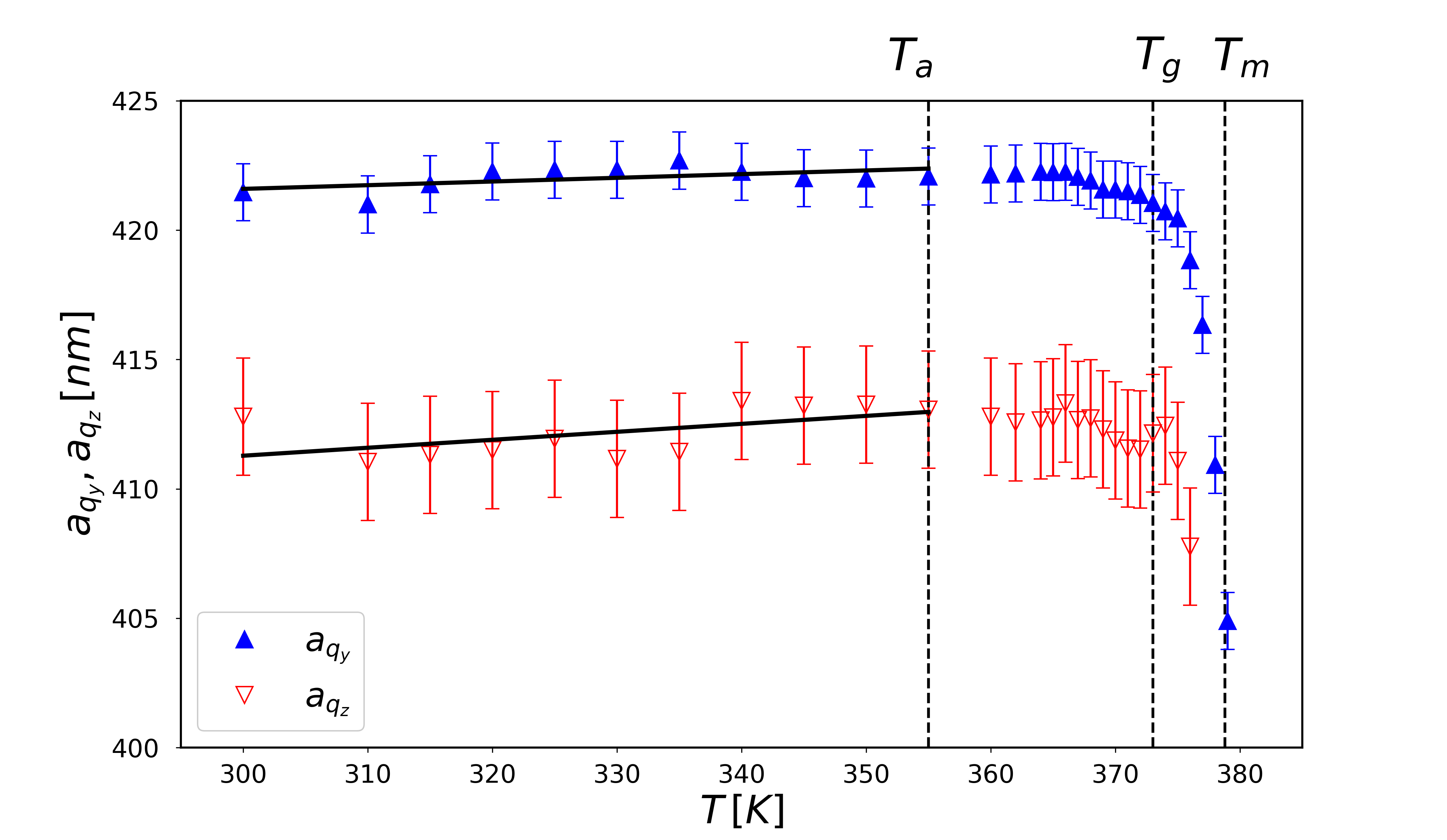}
	\caption{Temperature dependences of in-plane $a_{q_{y}}$ ($\vartriangle$) and out-of-plane $a_{q_{z}}$ ($\triangledown$) interparticle distances for the sample A. Linear fits are plotted by solid lines.}
	\label{img:Figure5}
\end{figure}

To unravel the behaviour of the colloidal crystals under dry sintering conditions we analysed GTSAXS data by evaluating the diffraction peak parameters, such as position, integrated intensity and widths as a function of temperature. The in-plane and out-of-plane directions in a colloidal crystal were probed by choosing the sets of (110) and (001) diffraction orders (Figure \ref{img:Figure2}). Thus obtained temperature dependencies of peak parameters for the sample A are shown in Figure \ref{img:Figure4} (results for the samples B and C are shown in Supplementary Figures \ref{img:S2}, \ref{img:S3}). At the initial stage of temperature raise up to $T_{a}$ the colloidal crystal undergoes thermal expansion while maintaining the long-range order. In this temperature range both in-plane and out-of-plane q-values of diffraction peaks decrease linearly. By converting q-values to interparticle distances we determined the thermal expansion coefficient of the PS crystal from a linear fit (Fig. \ref{img:Figure5}). The obtained in-plane and out-of-plane linear coefficients $\alpha_{L}$ of thermal expansion of PS of 6.4$\times$10$^{-5}$ $K^{-1}$ and 7.2$\times$10$^{-5}$ $K^{-1}$ are in a good agreement with literature data \cite{Beaucage} reporting the volumetric thermal expansion coefficient $\alpha_{V}$=3$\alpha_{L}$ for PS as 1.7 - 2.4$\times$10$^{-4}$ $K^{-1}$. Further temperature increase up to the glass  transition temperature of PS $T_{g}$ = 373 K results in decreasing of particle size due to the faceting of colloidal spheres. In a range from $T_{g}$ to the melting temperature $T_{m}$ the particle size decreases more sharply which is caused by the fusion process of PS particles. From the integrated intensity plot (Figure \ref{img:Figure4}b,e) we can directly determine the crystal melting temperature $T_{m}$, i.e. the temperature corresponding to the amorphisation of a colloidal crystal and the formation of a polymer film. $T_{m}$ values of 378 K, 381 K and 377 K have been obtained for the samples A, B and C, respectively. The observed increase of $T_{m}$ with an increase of the film thickness conforms with the results of earlier studies of glass transition in PS films \cite{Keddie}. 

\begin{figure}
	\centering
	\includegraphics[bb=0 0 2500 1500, width=\columnwidth]{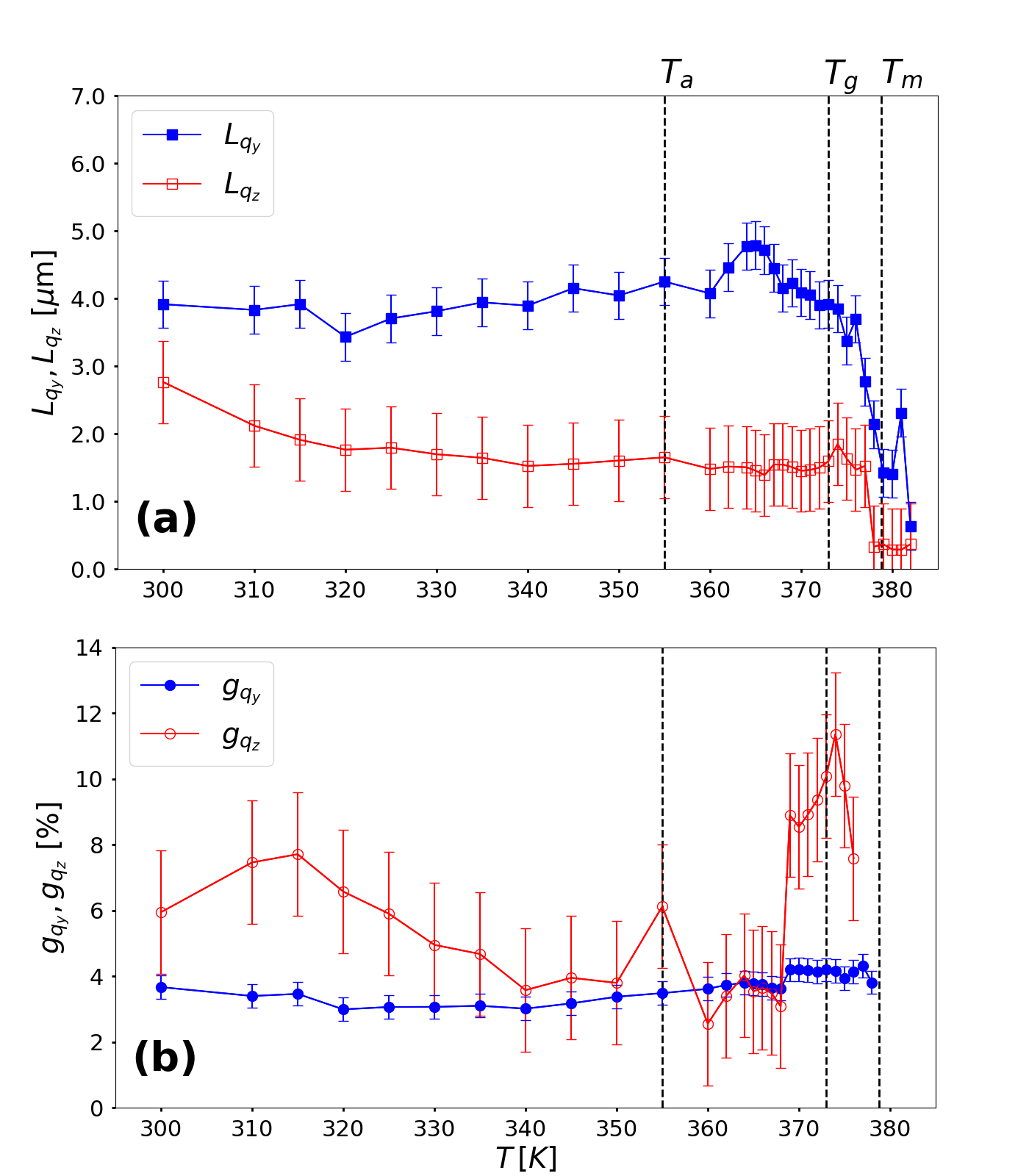}
	\caption{Temperature dependences of (a) CSD sizes and (b) lattice deformation parameters for in-plane (blue filled dots) and out-of-plane (red open dots) directions of the sample A.}
	\label{img:Figure6}
\end{figure}

Temperature dependences of peak widths $W_{q_{y,z}}$ were fitted using Williamson-Hall (WH) formula \cite{Sulyanova} :
$$W_{q_{y,z}}(T)^2=(2\pi/L_{q_{y,z}}(T))^2+ (g_{q_{y,z}}(T)q)^2,$$
where $L_{q_{y,z}}$ are coherent scattering domain (CSD) sizes and $g_{q_{y,z}}$ represent lattice deformation parameters. The temperature dependences of CSD sizes and lattice deformation parameters obtained from the fitting analysis are plotted in Figure \ref{img:Figure6} for the sample A. The CSD size values of 4 $\mu$m and 3 $\mu$m and lattice deformation values of 4\% and 6\% for in-plane and out-of-plane directions, respectively, mainly remain constant in a range from RT to annealing temperature. At $T$ = 365 K the in-plane CSD size exhibits a peak which coincides with a minimum of out-of-plane $g$ parameter, indicating the initiation of an intermediate ordered phase in the colloidal crystal due to thermal annealing. Even more pronounced effect of the in-plane CSD size enhancement was observed for the sample C (Supplementary Figure \ref{img:S4}), which allows us to conclude that the out-of-plane strain release leads to an enhancement of in-plane ordering in a colloidal crystal due to thermal annealing. With further raising of temperature up to $T_{g}$ the $L$ values decrease while the $g$ values increase, thus indicating the accommodation of lattice strains due to the fusion of colloidal particles.

\begin{figure}
	\centering
	\includegraphics[ bb=0 0 2000 600, width=\columnwidth ]{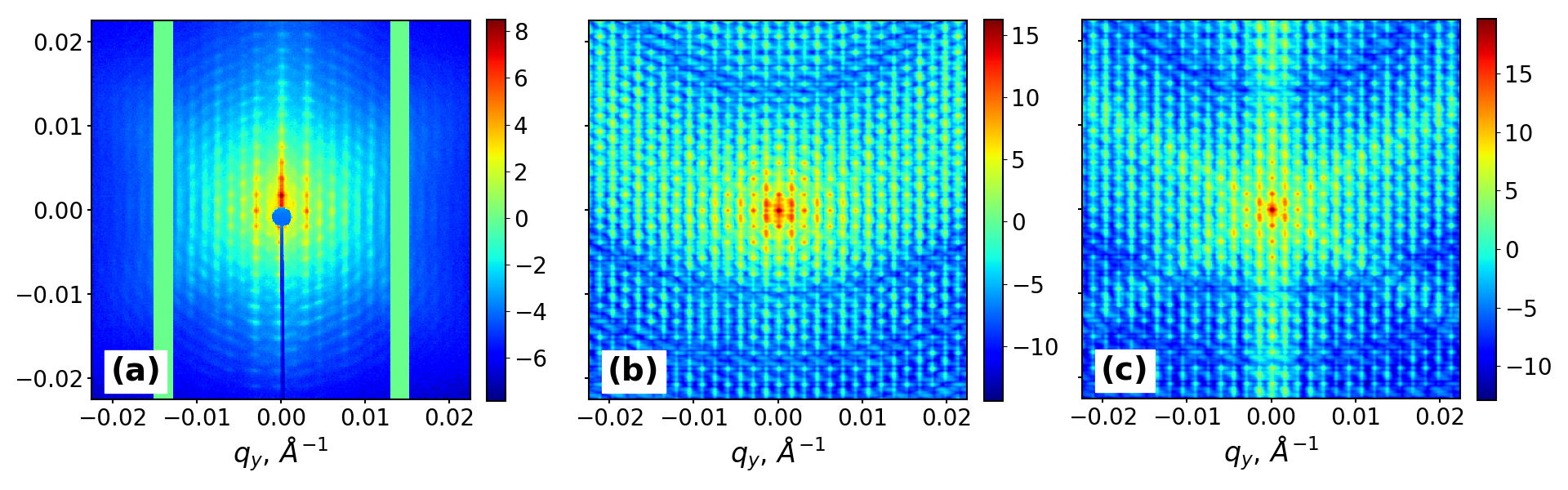}
	\caption{GTSAXS patterns for the sample A at $T$ = 376 K: (a) experiment, (b) simulation using the scattering function of a single sphere and (c) simulation using the scattering functions of a sphere and a rhombic dodecahedron.}
	\label{img:Figure7}
\end{figure}

At temperatures above $T_{a}$ the transformation of spherical particle to a faceted shape sets in, as indicated by an enhancement of anisotropic diffuse scattering pattern in a form of inclined streaks. Similar enhancement of diffuse scattering due to particle faceting has been observed previously in transmission SAXS geometry \cite{Zozulya, Sulyanova}. It is to note that the diffuse scattering enhancement observed in GTSAXS geometry is less pronounced than in the case of SAXS geometry because of the averaging of scattering signal in \mbox{GTSAXS} due to a larger beam footprint. We analysed the effect of particle shape transformation by GTSAXS simulations implementing the scattering function of a rhombic dodecahedron, which can be computed as a sum of scattering functions of a cube and six equilateral square pyramids attached to the faces of a cube \cite{Senesi}.  In Figure \ref{img:Figure7} the GTSAXS pattern measured for the sample A at $T$ = 376 K is compared to the simulated GTSAXS patterns, corresponding to the scattering function of a single sphere (Figure \ref{img:Figure7}b)  and a 1:1 mixture of spheres and rhombic dodecahedrons (Figure \ref{img:Figure7}c). As can be deduced from the scattering patterns and corresponding intensity profiles (Supplementary Figure \ref{img:S5}), the combination of spherical and dodecahedron shapes provides better description of the experiment as compared to the single sphere case. This result further implies that the shape transformation affects not all particles in a crystal, which is apparently caused by the presence of disordered domains. The transition of PS colloidal particles from spherical to a dodecahedron shape is also indicated by the decrease of lattice spacing at temperatures above $T_{g}$. As can be seen from Figure \ref{img:Figure5}, the in-plane lattice spacing decreases from 420 nm down to 405 nm, reaching the expected value $D_{rd}$ $\approx$ 0.96$D_{sp}$ for the lateral size of a rhombic dodecahedron of the same volume as an initial sphere of diameter $D_{sp}$ \cite{MyNote1}.

\begin{figure}
	\centering
	\includegraphics[ bb=0 0 2700 900, width=\columnwidth ]{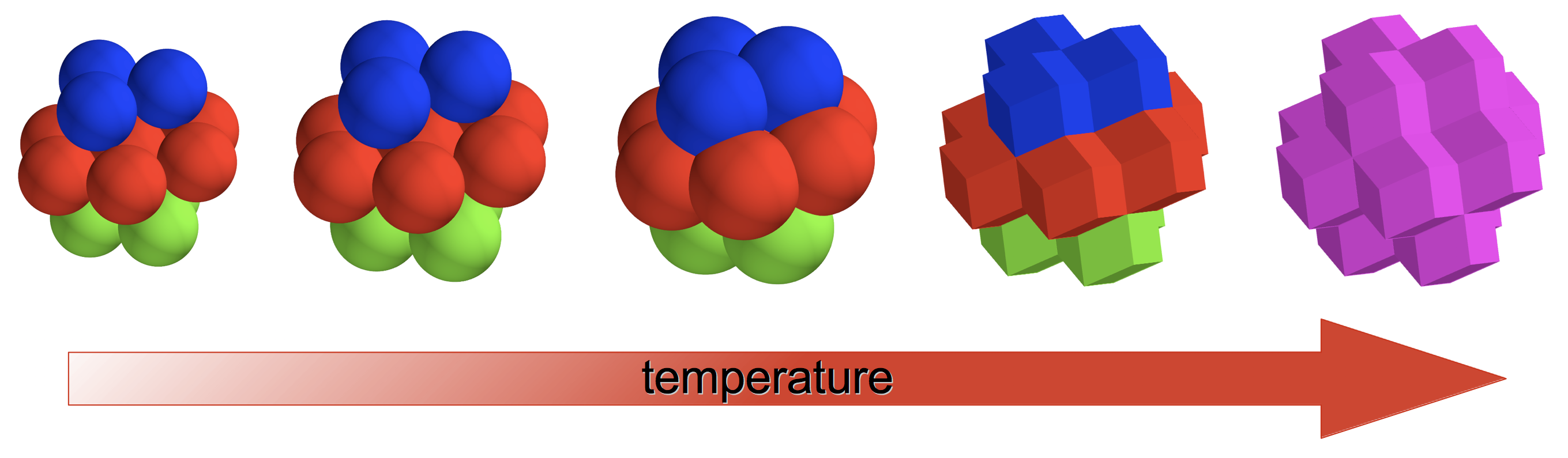}
	\caption{Schematic 3D model illustrating the structural rearrangement of \textit{fcc} closed-packed colloidal spheres in the process of dry sintering.}
	\label{img:Figure8}
\end{figure}

\begin{figure}
	\includegraphics[bb=-1500 0 8000 10000, width=\textwidth]{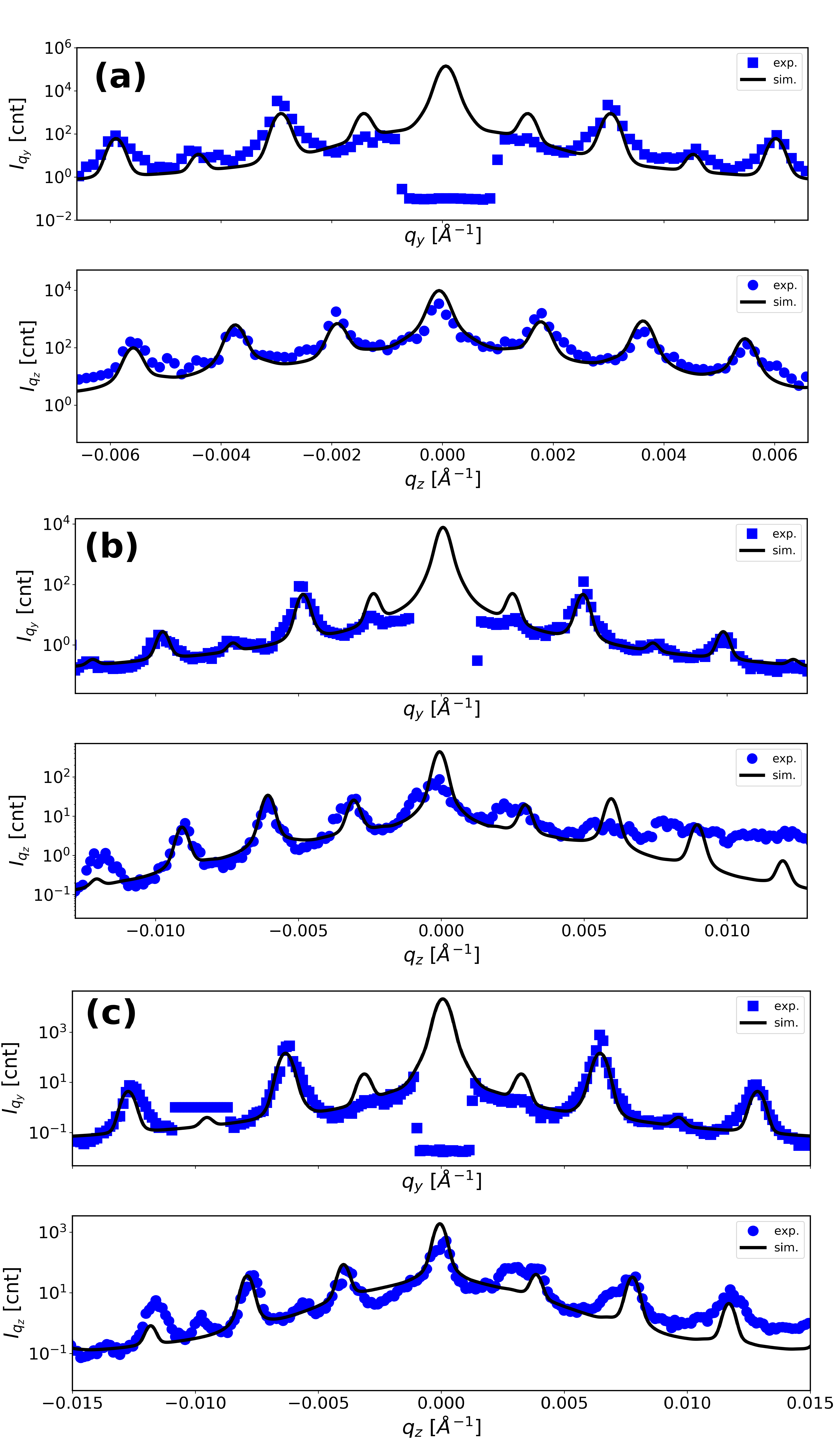}
	\caption{GTSAXS intensity profiles (dots) and fitting curves (lines) for the PS colloidal crystal samples (a) A, (b) B and (c) C at RT. For each sample the top and the bottom profiles refer to $q_{y}$- and $q_{z}$-directions (depicted as dashed lines in Figure \ref{img:Figure2} of the main text), respectively. $q_{y}$-profiles include diffraction peaks up to (220) order and  $q_{z}$-profiles include peaks up to (003) order. Structural parameters deduced from fitting curves are summarized in Table \ref{Table1} of the main text.}
	\label{img:S1}
\end{figure}

Structural evolution of a PS colloidal crystal in the process of dry sintering is visualised by a schematic 3D model in Figure \ref{img:Figure8}, introducing a close-packed \textit{fcc} arrangement of colloidal spheres with a central sphere surrounded by its 12 neighbouring spheres. The middle layer consisting of the central sphere and its 6 neighbouring spheres (shown in red) is sandwiched between the adjacent layers of 3 spheres each (shown in blue and green colours). Thermal expansion stage is followed by particle softening, which results in the particle shape transformation from a sphere to a rhombic dodecahedron. The shape transformation is followed by the particle fusion stage, where the boundaries between particles merge together due to the interdiffusion of polymer chains. 

\section{Conclusions}

In summary, the structural rearrangement of polymer colloidal crystals in the process of dry sintering has been revealed by \textit{in situ} GTSAXS measurements, which are sensitive to both in-plane and out-of-plane ordering. By simulations of GTSAXS patterns based on DWBA theory the structural characteristics such as particle size, number of layers, domain sizes, lattice disorder and mosaic spread have been determined at RT for the colloidal crystals made of spherical PS particles of different sizes. We observed the in-plane lattice dilatation of 1-2\% as compared to the out-of-plane direction in the PS colloidal crystals due to entropy-driven \textit{rhcp} arrangement of particle layers and surface tension effects.

We found that the PS colloidal crystals undergo several stages of structural evolution in the process of dry sintering: thermal expansion, particle shape transformation and crystal amorphisation. By analysing the peak positions as a function of temperature the linear coefficient of thermal expansion of polystyrene was determined, being in excellent agreement with literature data. At temperatures around glass transition temperature of PS we have observed characteristic enhancement of diffuse scattering indicating the particle transformation from spherical to polyhedron shape. The observed formation of polyhedral colloidal particles under dry sintering conditions in PS colloidal crystals has been supported by GTSAXS simulations using the particle shape functions of a sphere and a rhombic dodecahedron. 

At temperatures approaching the crystal melting temperature \textit{$T_{m} $} the softening of colloidal particles and interdiffusion of polymer chains lead to the blurring of interfaces between particles and the subsequent formation of an amorphous polymer film. We observed linear dependence of \textit{$T_{m} $} on the thickness of a colloidal crystal, in agreement with previous studies of glass transition in PS films. The observed decay of diffraction peak intensities and scattering domain sizes around \textit{$T_{m} $} implies that complete loss of ordering and formation of a polymer film occur within a narrow temperature interval of less than 3 K. 

Our study enables a novel approach to study the structural rearrangement of polymer colloidal crystals under dry sintering conditions. The revealed phase behaviour and intermediate states of colloidal particles in a crystal provide new insights to the self-assembly of soft matter nanostructures with on-demand optical properties. 

\section*{Conflicts of interest}
There are no conflicts to declare.

\section*{Acknowledgements}
Authors thank whole P10 team and especially M. Kampmann and D. Weschke for their excellent technical support of the experiment. We thank  M. Schroer for careful reading of the manuscript and valuable remarks. The research was carried out at PETRA III at DESY, a member of Helmholtz Association (HGF).


\bibliography{CC_GTSAXS_arXiv} 

\begin{mcitethebibliography}{59}
\bibitem[MyN()]{MyNote1}
The volume of a sphere $V_{sp}$=(4/3)$\pi$$R_{sp}^3$ has to be equal to the
  volume of a rhombic dodecahedron $V_{rd}$=(16$\sqrt{3}$/9)a$^3$, where the
  radius of a sphere $R_{sp}$=0.5$D_{sp}$ and $a$ is the edge of a rhombic
  dodecahedron. From this condition, the lateral size of a rhombic dodecahedron
  $D_{rd}$=$\sqrt{3}$a $\approx$ 0.96$D_{sp}$.\relax
\mciteBstWouldAddEndPunctfalse
\mciteSetBstMidEndSepPunct{\mcitedefaultmidpunct}
{}{\mcitedefaultseppunct}\relax
\EndOfBibitem
\end{mcitethebibliography}
\bibliographystyle{rsc} 

\beginsupplement
\vspace{5 cm}
\textbf{ELECTRONIC SUPPLEMENTARY INFORMATION}
\vspace{0.3cm}


\begin{figure}
	\includegraphics[bb=-1000 -1200 5500 5500, width=\textwidth ]{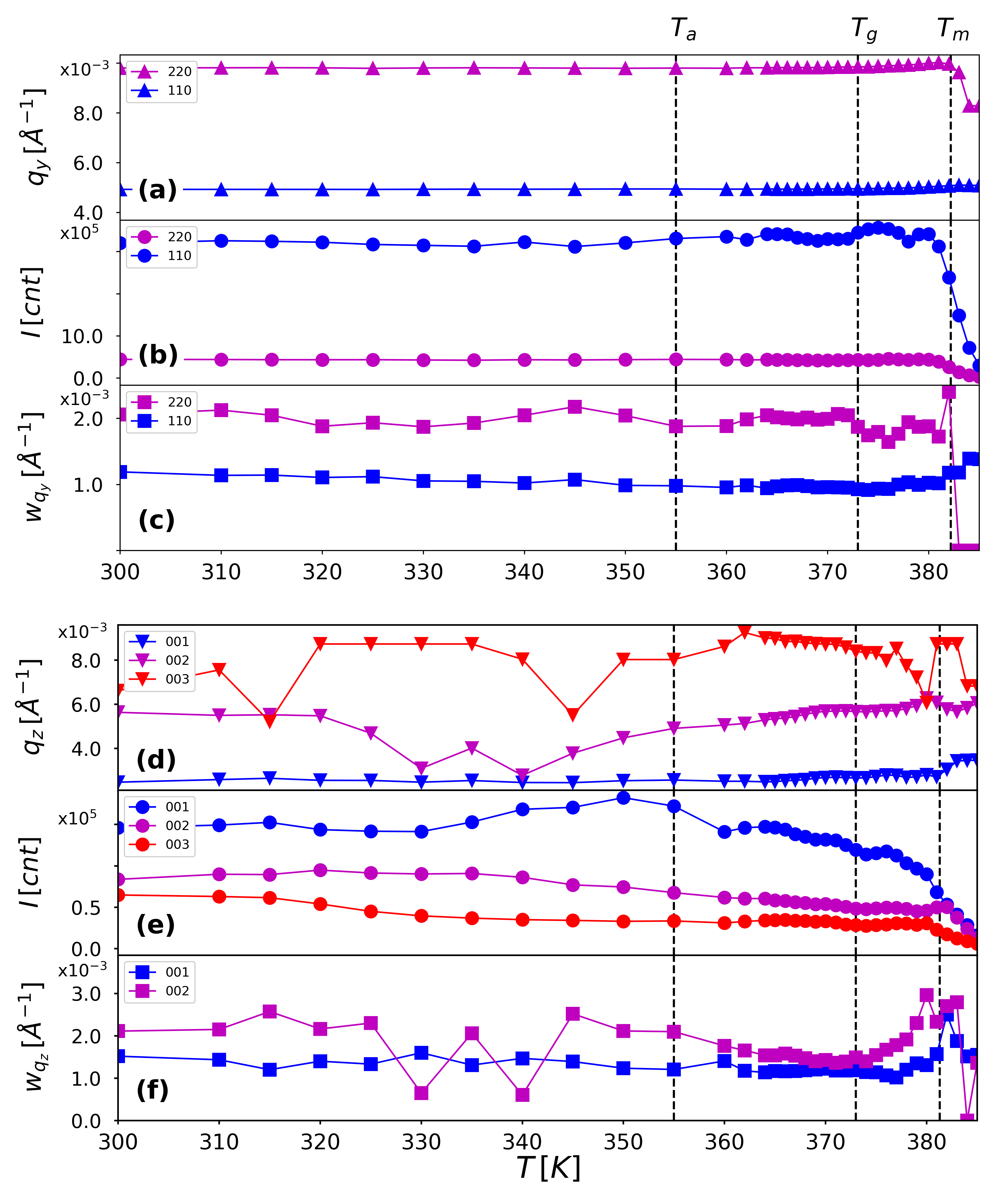}
	\caption{Temperature dependences of GTSAXS peak positions, integrated intensities and widths of the PS colloidal crystal sample B for (a-c) in-plane and (d-f) out-of-plane directions.}
	\label{img:S2}
\end{figure}

\begin{figure}
	\includegraphics[bb=-1000 -1200 5500 5500, width=\textwidth ]{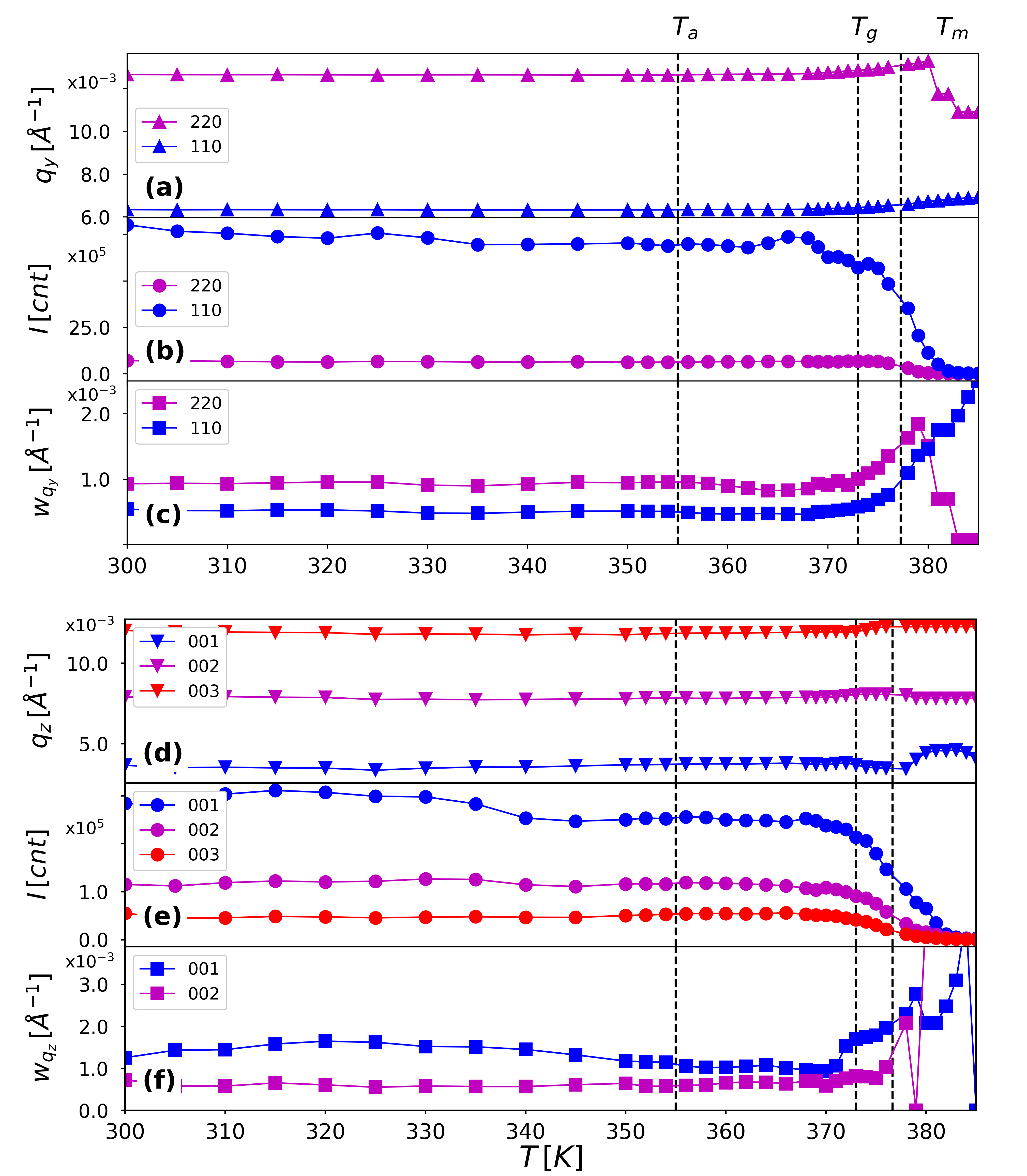}
	\caption{Temperature dependences of GTSAXS peak positions, integrated intensities and widths of the PS colloidal crystal sample C for (a-c) in-plane and (d-f) out-of-plane directions.}
	\label{img:S3}
\end{figure}

\begin{figure}
	\includegraphics[bb=-300 -360 1800 1800, width=\textwidth]{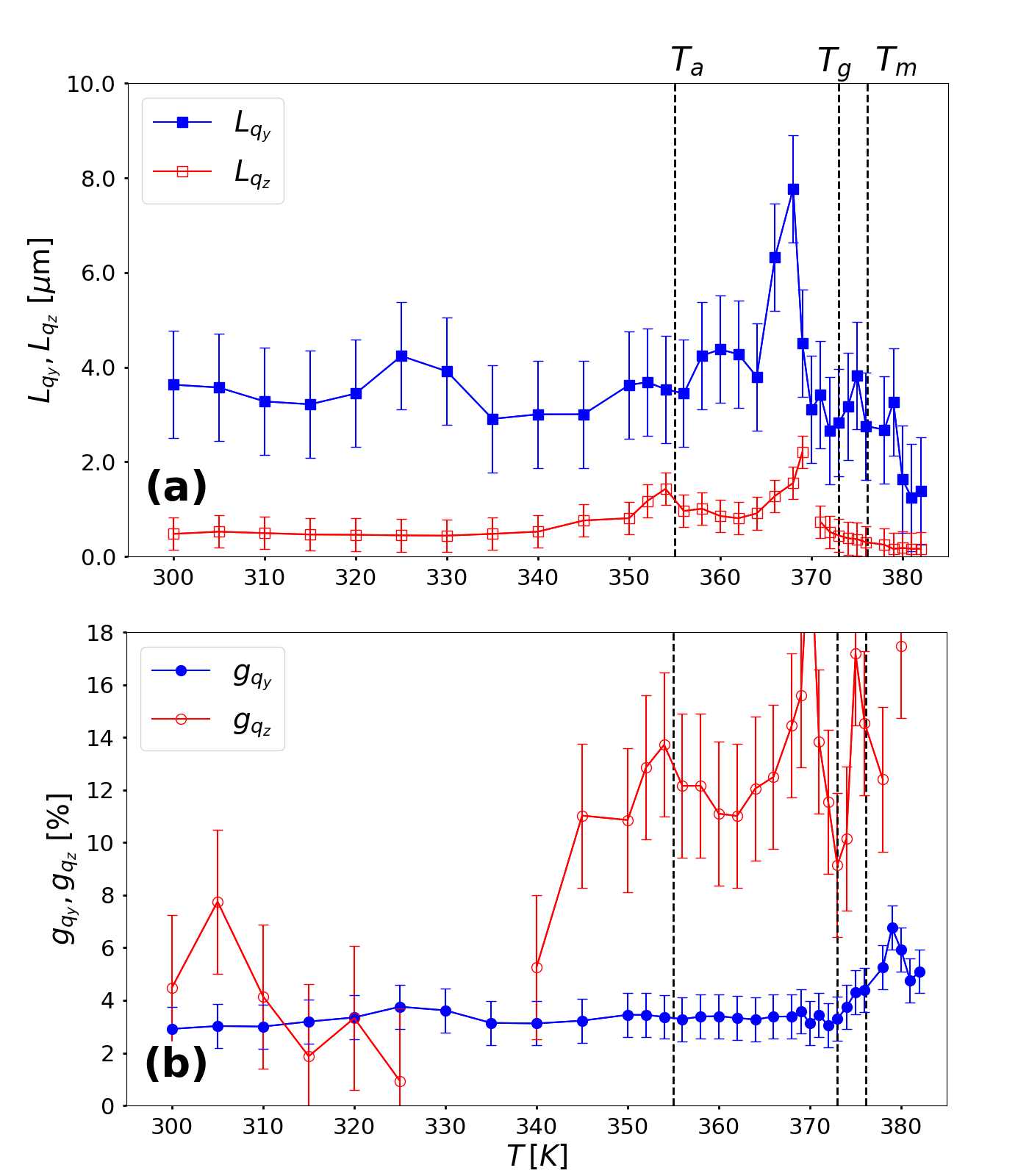}
	\caption{Temperature dependences of (a) CSD sizes and (b) lattice deformation parameters for in-plane (blue filled dots) and out-of-plane (red open dots) directions of the sample C.}
	\label{img:S4}
\end{figure}

\begin{figure}
	\includegraphics[bb=-100 -500 1200 800, width=\textwidth]{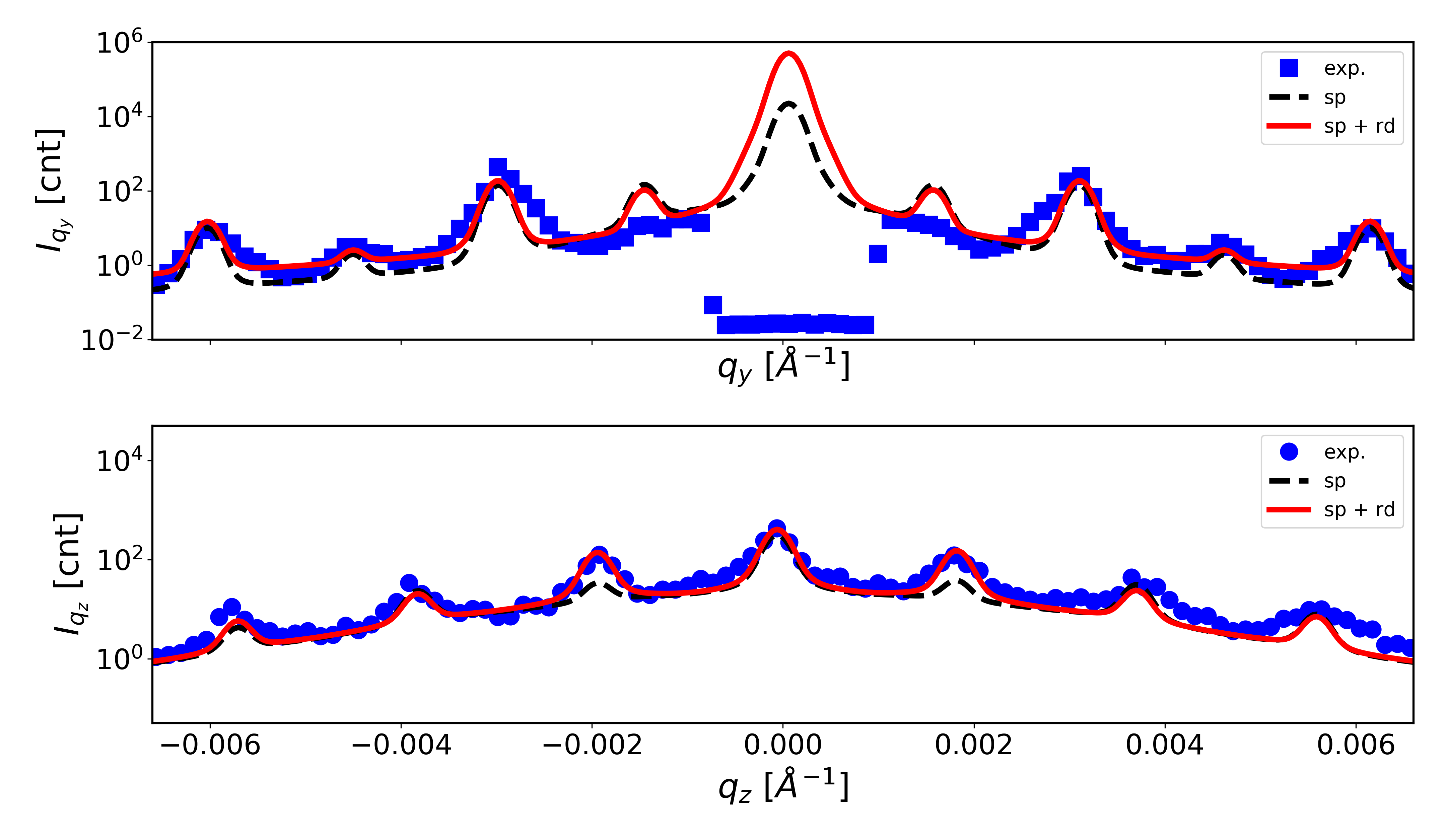}
	\caption{ Experimental (dots) and simulated (lines) GTSAXS intensity profiles for the PS colloidal crystal sample A at $T$ = 376 K. Calculated curves were obtained using the scattering functions of a single sphere (dashed line) and 1:1 mixture of spheres and rhombic dodecahedrons (solid line).}
	\label{img:S5}
\end{figure}

\end{document}